# Preliminary optical design of PANIC, a wide-field infrared camera for CAHA

M. C. Cárdenas[*a], J. Rodríguez Gómez[a], R. Lenzen[b], E. Sánchez-Blanco[a] and the International PANIC team[†]

[a]Instituto de Astrofísica de Andalucía (IAA-CSIC), P.O. Box 3004, E-18080 Granada, Spain
[b]Max Plank Institut für Astronomie (MPIA), Königstuhl 17, 69117 Heidelberg, Germany

## ABSTRACT

In this paper, we present the preliminary optical design of PANIC[‡] (PAnoramic Near Infrared camera for Calar Alto), a wide-field infrared imager for the Calar Alto 2.2 m telescope. The camera optical design is a folded single optical train that images the sky onto the focal plane with a plate scale of 0.45 arcsec per 18 μm pixel. A mosaic of four Hawaii 2RG of 2k x 2k made by Teledyne is used as detector and will give a field of view of 31.9 arcmin x 31.9 arcmin. This cryogenic instrument has been optimized for the Y, J, H and K bands. Special care has been taken in the selection of the standard IR materials used for the optics in order to maximize the instrument throughput and to include the z band. The main challenges of this design are: to produce a well defined internal pupil which allows reducing the thermal background by a cryogenic pupil stop; the correction of off-axis aberrations due to the large field available; the correction of chromatic aberration because of the wide spectral coverage; and the capability of introduction of narrow band filters (~1%) in the system minimizing the degradation in the filter passband without a collimated stage in the camera. We show the optomechanical error budget and compensation strategy that allows our as built design to met the performances from an optical point of view. Finally, we demonstrate the flexibility of the design showing the performances of PANIC at the CAHA 3.5m telescope.

## 1. INTRODUCTION

In 2005, the German-Spanish Observatory of Calar Alto (CAHA) changed its administrative status, and since then Germany and Spain have given financial support in equal shares for the development of new instrumentation for the observatory through their institutions Max-Planck-Gesellschaft (MPG) and Consejo Superior de Investigaciones Científicas (CSIC).

In the frame of this agreement, the idea of building a new infrared camera for the 2.2m telescope, as the first instrument of the consortium, was suggested by both astronomical communities, supported by the Scientific Advisory Committee and approved by the Executive Committee.

The optical design and expected performances of PANIC are presented hereafter.

The camera will have a Field of View (FOV) of 31.9'x31.9', with a diameter only one arc-minute smaller than the largest non-vignetted FOV that the 2.2 m telescope allows without field corrector. Several scientific cases require a plate scale of 0.45 arc-sec per pixel and therefore a mosaic of four detectors of 2kx2k pixels is envisaged. Others scientific projects, which require smaller plate scale, drive the study of the camera for the 3.5m telescope. In this case the FOV is 16.4'x16.4' with 0.23 arc-sec/pixel.

The schedule of the design and construction of the camera is following the next milestones: Kick-off October 2006,, PDR (Preliminary Design Review) successfully completed in November 2007, FDR (Final Design Review) for optics in July 2008 and in the third quarter of year for the other work packages, finally first light at telescope during 2011.

---

[*] conchi@iaa.es; www.iaa.es
[†] See http://www.iaa.es/PANIC/index.php/gb/panic_team
[‡] See http://www.iaa.es/PANIC/

## 2. INSTRUMENT PARAMETERS

The general requirements for PANIC from the start are:

• 2.2m telescope, Ritchey-Chrétien (RC) focus.

• Detector size 4096x4096 pixel.

• Spectral range Near Infrared (NIR), i.e. minimum YJHK.

• Image scale 0.45 arcsec/pixel.

These basic requirements have direct consequences on the design of PANIC. First of all, the instrument must not exceed the limits set by the telescope in size, weight and envelope at the RC focus of the CAHA 2.2 telescope (for more details see [1]). We have studied different alternatives for the optical design (e.g. [3] and [4]) to make the most of the RC focus capabilities (see the first seven rows in Table 3).

Finally, the optical system is a monobeam design, all refractive, being the only mirrors of the system the ones used for folding and packaging. The design has not been required to have an internal collimated beam. The optical design produces an internal pupil available for a Lyot stop at the telescope image pupil placed at the primary mirror.

While designing PANIC, several additional features were proposed which go beyond the basic requirements. The ones we followed up are:

• Extend the spectral range to 0.82 μm, so PANIC will cover all spectral bands from the z to K. The z-band has been included for convenience of the observers, in order to allow z-band observations to complement NIR observations without changing instrumentation or waiting for another instrument to be mounted. The applications of PANIC, however, are in the NIR.

• The use of narrow band (bandwidth = 1% of central wavelength) filters. This requires that the angle of incidence of the beam does not exceed a value of 10º. Our optical design takes this into account.

• The possibility to move occasionally PANIC to the 3.5 m telescope of CAHA, witch represents a factor 2 in scale. This image scale will allow higher spatial resolution and will be very useful under good seeing conditions which prevail frequently since the median seeing at Calar Alto is 0.89 arcsec in the R band which corresponds to 0.65 arcsec in the K band.Table 1 summarizes the general specifications for PANIC established/imposed by the science goals and the technical requirements that derivate of the operational conditions and design choices.

Table 1: Summary of the PANIC general specifications.

| Focal Station | Cassegrain 2.2 m |
|---|---|
| FOV | 30' x 30' |
| Pixel scale | 0.45 arcsec/pixel |
| Direct Imaging | Over the whole FOV |
| Image Quality | EE80 ≤ 2 pixels = 0.9 arcsec = 36 μm |
| Distortion | ≤ 1.5 % |
| Pupil image available | Cold stop |
| Wavelength range | 0.95– 2.45 μm with IQ<br>0.82-0.95 μm able to transmit |
| IR Detector | 4 K x 4 K |
| Gap between detectors | Minimum |
| Operating temperature | 80 K (liquid nitrogen) |
| Filters | Broad band: zYJHK<br>Narrow band: ~1%[§] |
| System focusing mechanism | Telescope S2 |
| Performance evaluation at | Cassegrain 3.5 m |

---

[§] This value is calculated: Full Width at Half Maximum (FWHM) divided by the filter central wavelength, expressed in %.

## 3. OPTICAL DESIGN

As mentioned before, the camera optical design is a single optical train that images the sky onto the focal plane at an optical speed of f/3.74, with a plate scale of 0.45"/pix. Fig. 1 shows the optical straight layout which is 1890 mm long, as well the folded solution.

The camera consists of one field lens, L1, close to the RC telescope focus; and two separate groups of lenses, one, from L2 to L5, before the cold stop, and another, from L6 to L9 after the cold stop mask. Due to the mechanical constrains in length and weight it has been searched alternatives to make the system more compact and finally the packaging solution adopted introduces three folding flat mirrors in the optical path between L1 and L2. From the optical performance point of view this packaging proposed has not effect. The mirrors positions have been fixed for an optimum separation, with no possible interference and vignetting in order to reduce the cold volume of the system.

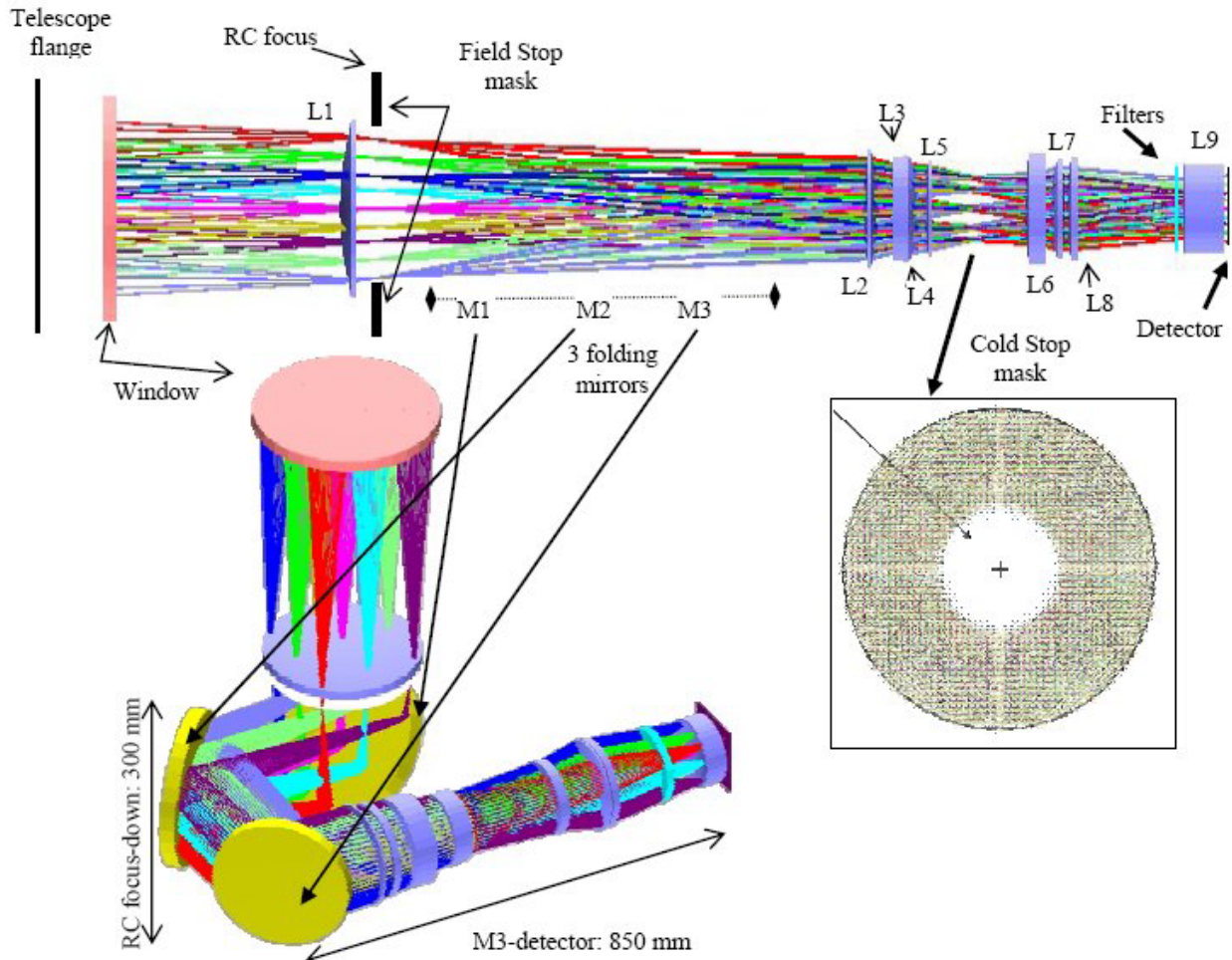

Fig. 1. PANIC optics layout, including unfolded layout.

The PANIC optics design has been modeled for cryogenic temperatures and vacuum using a glass catalogue at 80 K produced for that purpose. The optical prescription of the system is listed in Table 2. The curvature radius and thicknesses of the lenses are given at 80 K, working temperature of PANIC. For manufacturing and assembly, those parameters have to be replaced by warm parameters, using thermal expansion coefficients (CTE) calculated from room temperature to 80 K. The models for calculation of the refraction index variation with temperature and the CTE are based on [5], [6], [7], [8] and [9].

Table 2: Prescriptions data of the optical system at its nominal design temperature.

| Element | Curvature radius (mm) at 80 K | | Center Thickness (mm) at 80 K | Material | Full Aperture diameter (mm)r |
|---|---|---|---|---|---|
| | Front face | Rear face | | | |
| Cryostat window | Infinity | Infinity | 20.0 | IR Fused Silica | 330.00 |
| L1 | 430.1 | Infinity | 28.0 | IR Fused Silica | 255.00 |
| Field stop | Infinity | -- | | | |
| M1 | Plane | -- | 27.5 (TBD) | Fused silica | 282.00 |
| M2 | Plane | -- | 24.8 (TBD) | Fused silica | 255.00 |
| M3 | Plane | -- | 22.0 (TBD) | Fused silica | 229.00 |
| L2 | 415.60 | -267.0 | 27.1 | CaF2 | 170.00 |
| L3 | -176.80 | -423.0 | 8.0 | S-FTM16 | 152.00 |
| L4 | -141.48 | -135.77 | 9.6 | IR Fused Silica | 154.00 |
| L5 | 277.00 | 4765 | 14.4 | BaF2 | 142.00 |
| Cold stops | Infinity | -- | -- | -- | -- |
| L6 | 485.90 | 143.40 | 8.0 | S-FTM16 | 160.00 |
| L7 | 170.00 | -831.00 | 23.6 | BaF2 | 143.00 |
| L8 | 310.20 | -10547 | 21.7 | IR Fused Silica | 150.00 |
| Filter | Infinity | Infinity | 8 (TBC) | N-BK7 | 125.00 |
| L9 | -115.90 | 261.40 | 38.6 | IR Fused Silica | 130.00 |

As the filters will be placed in the convergent beam, they have been simulated by inserting a plate of N-BK7 with a thickness of 8 mm between the L8 and L9. The chief rays for different fields at the filter location were constrained to be parallel, thus removing most of the field dependence of any wavelength shift due to the change in incidence angle with field over the filters.

For interference filters, because of the non collimated condition, the expected filter performance will suffer a broadening of the apparent band pass, a depression of transmittance values and a shift to shorter wavelengths. For broadband filters the effect is negligible. For narrowband filters we have to calculate carefully this effect and determine the incidence angle which is a flux-weighted mean of the final converging beam to specify to the manufacturers the filter to operate at that angle. Due to the constrains imposed during the optical design we do not expect any problem with this, even for %1 narrowband filters.

The FOV has been sampled and optimized from the centre to the external field in a radial configuration following the equal area rule to cover the complete detector surface and the spectral band from 0.95 to 2.45 μm. The Table 3, first column, lists a summary of the characteristics that describe the optical performance of PANIC in the 2.2 m telescope.

The image quality of the instrument is specified in terms of the 80 % Ensquared Energy (EE80). For simplicity, it has been presented, in Fig. 2, left, only the polychromatic EE using the highest value obtained in the complete FOV. All the bands are in requirements (EE80 ≤ 2 pixels). In the X axis is the half side length square of EE and the Y axis represents the fraction of energy enclosed, where it is indicated with a horizontal line the 80%. The diffraction limit of the system is shown in dark. For simplicity, as well, it has been presented only the polychromatic spot diagram in Fig. 2, right. This figure shows the geometrical structure of the image at all points of the field for all the wavelengths considered. The squared boxes indicate the dimension of two pixels in the focal plane (36 μm), and the Airy disk for this configuration is indicated with the dark circle. Better figures are obtained when the system is refocused in the photometric bands. Although the system has not been optimized for z band, the results give us optical quality in this extreme band due to the selection of materials. So finally the system complies the same requirements in that photometric band as the others.

The distortion has been calculated in % with respect to the FOV centre which does not have distortion. In the transmission calculations the three folding mirrors has been considered with a gold coating.

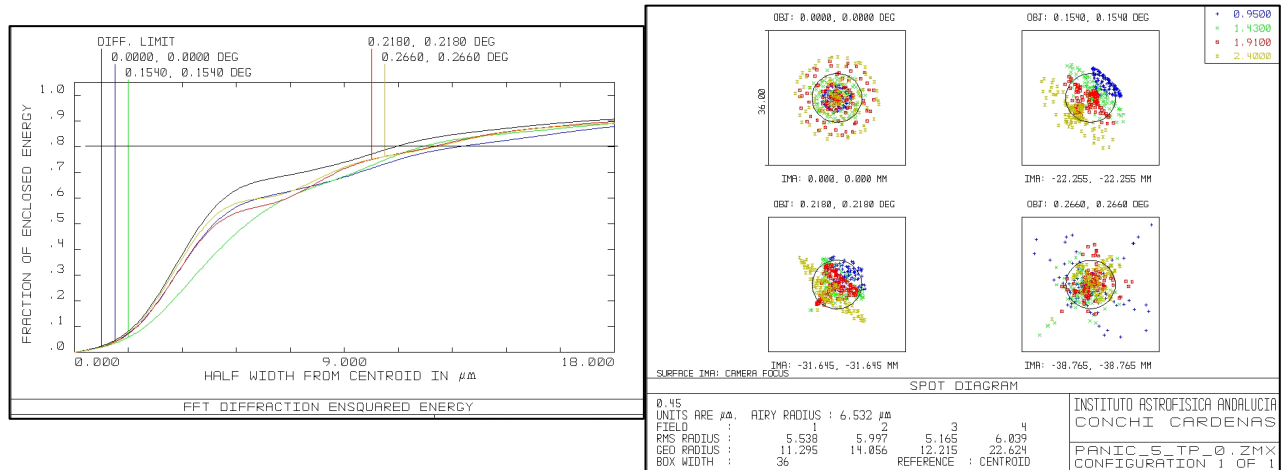

Fig. 2. Polychromatic EE and spot diagram for PANIC in the 2.2 m telescope. Boxes are 2x2 pixels.

Table 3: General capabilities of the 2.2 m and 3.5 m RC foci and summary of the PANIC general specifications.

| CAHA Telescope | @ 2.2 m RC focus | @ 3.5 m RC focus |
|---|---|---|
| Optics | Ritchey-Chrétien | Ritchey-Chrétien |
| Aperture, ∅ S1 | 2.2 m | 3.5 m |
| Focal ratio | f/8 | f/10 |
| HFOV with no vignetting | 0.276º | 0.245º |
| ∅ Cassegrain focus | 33' = 170 mm | 29.5' = 300 mm |
| Scale at Cass. focus | 11.7 "/mm | 5.89 "/mm |
| PANIC performances | | |
| Direct imaging | Over the whole FOV | Idem |
| FOV | 31.9' x 31.9' | 16.4' x 16.4' |
| Scale at detector | 0.45 "/px | 0.23 "/px |
| Pupil image mechanism | Mechanically available, Optimized for 2.2 m | Mechanically available, Optimized for 3.5 m |
| Pupil image quality | < 2% loss in flux all bands | Idem |
| Wavelength range | Optimized: 0.95 – 2.5 μm  Good transmission from 0.8 μm | Idem |
| Image Quality, EE80 | 1.5 pix.= 25.8 μm= 0.65" max. (≤ 2 pixels=36μm=0.8") | 2.1 pix.= 36.9 μm = 0.47" max (≤ 3 pixels=54μm=0.75") |
| Distortion | < 1.4 % max. (corner) | Idem |
| Transmission | ~ 69.6 % (window+9 lenses+3gold mirrors) | Idem |
| IR Detector | 4 K x 4 K | Idem |
| Operating temperature | 80 K | Idem |
| Gap between detectors | 167 pixels (minimum) | Idem |
| Filters | Broad band: zYJHK  Narrow band ~1% | Idem |

## 3.1 Stray Light strategy, Field Stop and Cold Stop

In order to minimize the stray light, the optical design of PANIC is baffled with the two naturally stops, a field and a pupil stop. All the lenses and mirrors have been over dimensioned to avoid stray light coming from the lens edges. The contribution due to ghosts has been minimized introducing several baffles in the optical path and no optical element with diamond turned surfaces (i.e. aspheric surfaces) is used. Finally the folded mirrors are gold coated to reduce imaging errors and scattered light.

In addition, the opto-mechanical design of PANIC uses a light tight optical labyrinth in the optical between the optical assemblies. The whole system is encapsulated to minimize stray light effects [1].

The Field stop is placed at the position of the RC focal plane, as shown in Fig. 1, between L1 and M1. This aperture is usually located at an image to limit and define the FOV without adding radiating flux from warm surfaces, which is critical in the K band, and no vignetting. This provides a good shielding from off-axis sources of light that would be outside the desired FOV.

In PANIC the Field Stop mask needed has been calculated for the 2.2 m and 3.5 m telescopes. The free opening proposed is square shape with the same orientation as the detector. The optimal positions of the field masks in axial direction, from the rear surface of L1 and its dimension, are optimized for the 2.2 m and these dimensions are coincident with the field stop needed for the 3.5 m. Therefore, only one field stop is needed to work in both telescopes.

The main stray light control feature in the optical design of a near infrared camera is its cold stop at the pupil image to reduce the thermal background, especially, in the K band. The Cold stop is used to suppress undesirable light that could reach the detector; it prevents the detector from seeing anything but the science beam path with the imaged scene, especially the warm interior of the system. In PANIC the entrance pupil has been placed at the telescope primary mirror, S1, which gives the maximum light collecting power and a good image of the secondary reflected in the primary.

The PANIC optical design provides a mechanically accessible pupil image with a good image quality of the S1 in the middle of the optical track as Fig. 1 shows. To achieve maximum background suppression and avoid loosing minimum flux in K band, we have proposed a mask with an outer hole, which corresponds to the re-imaging S1 diameter, and an inner mask, which corresponds to the S2 obstruction. The maximum degradation in the pupil re-imaging diameter is lower than 3% which is less than a 10% loss in flux for K-band. The optimal position and size calculated of this cold stop depends on the telescope in which PANIC will work, so both masks are mounted in a wheel to place them properly. During assembling, integration and verification (AIV) of the instrument it is planed to use a pupil imager located in one position of the first filter wheel. This pupil imager is composed by one lens and it is optimized to re-image the cold stop of the 2.2 m telescope on the detector. The tolerances calculated for both cold stops give us the possibility to use the same pupil imager to re-imagine the cold stop of the 3.5 m telescope too.

Secondary stray light sources (i.e. surface roughness modeled by BRDF) were not considered in the analysis as standard polishing techniques will be used in the manufacture. Otherwise it would involve an unjustified amount of time in the use of models and material finishing and scattering that we consider that it is not needed for the current performance goals and the time given to finish the project.

## 3.2 Ghost analysis and Baffling

The NSC (Non-Sequential Components) ghost analysis done for PANIC confirms the complete fulfillment of the requirements regarding the ghost radiance ratio to the nominal source as well as the minimum size. A quantitative ghost analysis has been done to identify the worst undesired paths. The proposed baffling strategy, resumed in Fig. 3, has been iterated with the mechanical design regarding final apertures and positions. These improvements will work on secondary paths that are low level sources by its nature. The higher results in intensity ratio between a ghost image and its source are under $1 \cdot 10^{-5}$ in all the cases that is well within specifications, smaller than $1 \cdot 10^{-4}$. Moreover, the smallest ghost structure is over 10" diameter, within requirement too. The contribution in intensity is insignificant, so the impact of the ghosts in the total PSF (Point Spread Function) of the system is negligible.

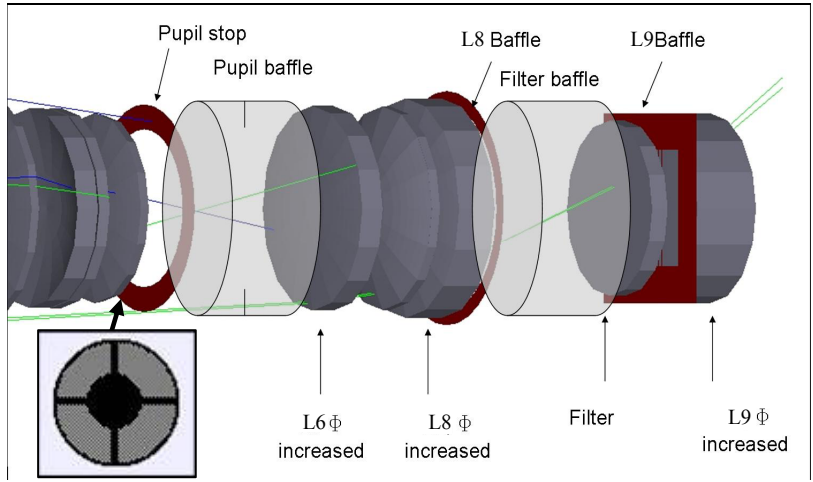

Fig. 3. Baffling proposal layout after the stray light analysis. L6, L8 and L9 diameters should be increased to avoid direct viewing of the lens walls.

## 4. PANIC AT THE 3.5 M TELESCOPE

PANIC working in the 3.5 m telescope provides a smaller pixel scale of 0.23"/pix., 16.4' square FOV and an image quality EE80 less than 0.75" (3 pixels) over the full FOV for each of the broad bands, including z band as has been required for the science cases. All the bands are within requirements (EE80 ≤ 3 pixels=54μm=0.75"). The Table 3, second column, lists a summary of the characteristics that describe the optical performance of PANIC at the 3.5 m telescope. The first seven rows show the RC focus capabilities of this telescope.

To have a better view, it has been presented, in Fig. 4, left, only the polychromatic EE using the greater value obtained in the complete FOV. For simplicity, as well, it has been presented only the polychromatic spot diagram in Fig. 4, right. The squared boxes indicate the dimension of two pixels in the focal plane (36 μm), and the Airy disk for this configuration is indicated with the dark circle. Again, better figures are obtained when the system is refocused in the photometric bands.

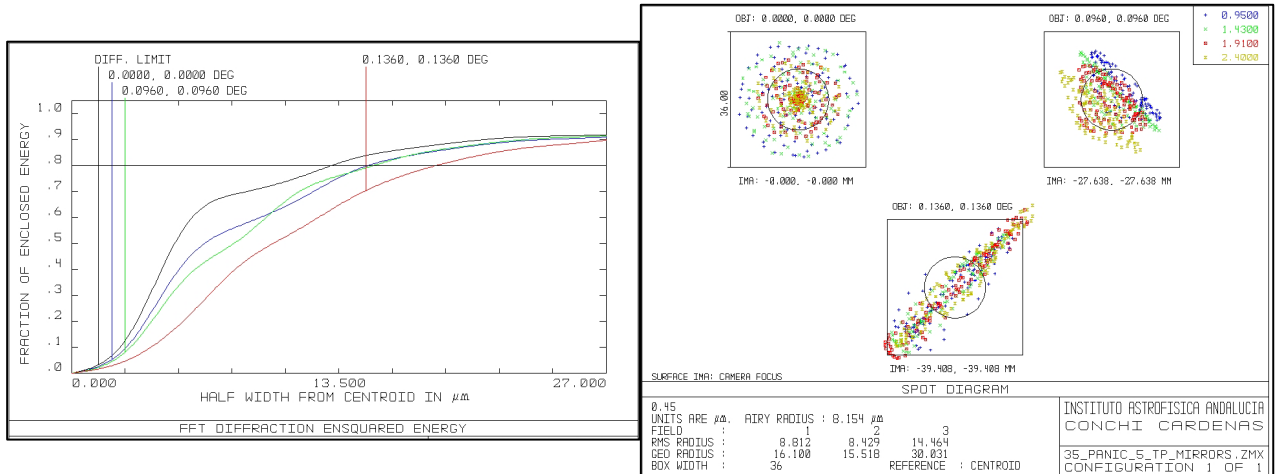

Fig. 4. Polychromatic EE and spot diagram for PANIC in the 3.5 m telescope.

## 5. TOLERANCE ANALYSIS AND AIV

A preliminary study of the tolerances for PANIC has been performed regarding image quality and image stability. The results are the tolerances needed to be defined for the optical manufacturing, position accuracy during assembly and stability during operation. The image quality nominal criterion to accept the degraded system is the 'half EE80' ≤ 2px=18μm for PANIC @ 2.2 m. The system has been evaluated in terms of the rms spot radius at five fields (FOV centre and 4 external situated at 90% of the FOV corner) and in three wavelengths (to cover the complete spectral range). The fabrication and alignment tolerances have been calculated separately, since we have decided to introduce the melt data (refraction indexes of the glass blanks) in the optics model and also the final dimensions of the manufactured elements to relax somewhat the alignment tolerances. Once the system is cooled, the only available adjust the instrument is by refocusing the telescope (using the S2) although for integration a detector adjustment in position and tilt is possible. The distance between S2 and the camera has been used as a compensator during tolerancing. The tilt in the detector will be required to compensate for the angle introduced when the decentering compensator are used.

For the element fabrication tolerances have been tolerated: the radius of curvature (ROC) of the two surfaces, front and rear (in the case of flat surfaces it has been tolerated the flatness in fringes); the thickness of the element (except for the mirrors); and the wedge. First results reveal that there are two distances which need compensation due to manufacturing errors: L2-L3 distance and L6-L7 distance. Both compensation will be in the order of ±0.5 mm. These distances will be done after the factory report of the as-built singlets, such as thicknesses, radii, wedges and lens distances are measured. A new optimization is then carried out and final values of these compensators are evaluated and fixed. Thus this compensator will only compensate for symmetrical aberrations. With these considerations the fabrication tolerances are in the precision quality level.

For alignment tolerances of the elements and sub-systems has been considered the grouping shown in Table 4 according with the mechanical design. The parameters tolerated are: the position in the axial direction, and the decenter and tilt in X and Y (being X and Y contained in the plane perpendicular to the Z axis).

Table 4: PANIC optical groups, including the FPA (Focal Plane Array). (LM, Lens Mount; L, Lens; M, Mirror)

| Optical element | L1 | M1 | M2 | M3 | L2 | L3 | L4 | L5 | Cold stop | L6 | L7 | L8 | L9 | FPA |
|---|---|---|---|---|---|---|---|---|---|---|---|---|---|---|
| Groups | LM 1 | mirror structure | | | LM 2a | | | | LM 2b | LM 3 | | | LM 4 | |
| | optics mount 1 | | | | | | | | | optics mount 2 | | | | |
| | complete optics | | | | | | | | | | | | | |

The first results obtained showed some elements with very tight tolerances, both in position and tilt, lower than 20 μm in decenter and 40" in tilt. Therefore, it has been decided to establish some compensators to relax the critical values as much as possible. They are: L2 decenter and L7 decenter. The analysis reveals that for L2 it is necessary a range for compensation of 400 μm in X and 300 μm in Y, as well as 250 μm in X and 350 μm in Y for L7. Those elements will be adjusted in decenter while placing an interferometer to cancel the non-symmetrical aberrations due to lens wedges and mounting tilts. These compensators have the effect of correcting non-symmetrical aberrations.

Summing up, the values for the integration tolerances after the compensators are implemented include each element, the different barrels and the whole instrument, nested as Table 4 shows. The smallest tilt values are in the order of 1 arcmin for elements and some barrels. Regarding the decenter tolerances, they result in a standard quality level, relaxed to 100 μm, as well as the position.

All of these results feed the opto-mechanical and alignment strategy of the instrument. In the following phase of PANIC a complete image quality error budget will be developed, to include thermal errors and image stability as well as its effect on tolerances.

A preliminary optical AIV plan has been made for the PANIC instrument covering engineering, tests regarding the optics. This AIV plan describes the procedures and equipment required for integration of the instrument and the verification tests since the integration in subsystem and system level will be in-house tasks. It is known how difficult the mounting of optical components in a cryogenic instrument is, due to the different coefficients of thermal expansion (CTE) of the optical elements and their mounts. The solution adopted in PANIC for the optical element is described in detail in [1] and [2].

It has been divided in three main categories related to the optical integration process from components manufacturing and tests, barrel integration (subassemblies) and tests, and system integration and final engineering tests. To design this plan it is necessary to identify the adjustments and compensators which come from the tolerance analysis. The different tasks and tests regarding each integration stage are described at each level (components, subsystem or system).

The optical AIV process will have two independent responsibilities. The optical elements manufacture will be accepted at the optical shop as individual elements and the integration of these lenses in the barrels and in the full instrument will be done by the PANIC team. The rationale behind the integration process is to test the functionality and performance of the different pieces at each assembling step as these are being installed. In that sense the system integration and verification should not display any fault at the subsystem or component level allowing a quick engineering and science verification. The barrels with decentering compensator will be assembled with an interferometric adjustment, and during the integration the compensator in distance will be adjusted. For Barrel 1, which does not have adjustments proposed, the alignment will be verified. All the sub-barrels will be cryogenically verified. Finally, the whole instrument will be assembled and tested, as we do not expect to need further adjustments than the mounting tolerances, the only adjustment to be done is the one for the detector, in position and tilt.

## CONCLUSIONS

The nominal optical design meets the desired performance criteria and contains margin to be applied to fabrication and alignment tolerances. To achieve this, a specific control plan during integration phases will be considered. A deep study of the tolerances and quality compensators, environmental change, and stray light has been done. The Optical design is currently being iterated with potential lens manufacturers and the mechanical designers.

The design contains only spherical surfaces (i.e. no conic or aspheric surfaces) and special care has been taken in the selection of lens materials not using high index refraction materials in order to include all the photometric bands, even the z band, in the system, avoiding some critical materials. The correction of off-axis aberrations due to the wide-field available, the correction of chromatic aberration due to the wide spectral coverage, the introduction of narrow band filters (~1%) in the system minimizing the degradation in the filter pass-band have been achieved with this optical design. An important point is the production of the internal cold stop with good optical quality which reduces the background in K band considerably. The feasibility of PANIC to work in both telescopes, the 2.2 m and the 3.5 m, has been presented.